\documentclass[preprint,prl,aps,amsfonts,amssymb,amsmath,showpacs]{revtex4}
\usepackage{graphicx}% Include figure files
\usepackage{dcolumn}% Align table columns on decimal point

\graphicspath{{figures/}}
\setkeys{Gin}{width=\linewidth}

\begin{document}
\title{Coherent tunnelling across a quantum point contact in the quantum Hall regime}

\author{F. Martins$^{1*}$, S. Faniel$^{1,2}$, B. Rosenow$^{3}$, H. Sellier$^{4}$, S. Huant$^{4}$, M. G. Pala$^{5}$, L. Desplanque$^{6}$,\\ X. Wallart$^{6}$, V. Bayot$^{1,4}$ and B. Hackens$^{1*}$\\}

\affiliation{ $^{1}$ IMCN/NAPS, Universit\'e catholique de Louvain, Louvain-la-Neuve B-1348, Belgium\\
$^{2}$ ICTEAM/ELEN, Universit\'e catholique de Louvain, Louvain-la-Neuve B-1348, Belgium\\
$^{3}$ Institute for Theoretical Physics, Leipzig University, Leipzig D-04009, Germany\\
$^{4}$ Institut N\'eel, CNRS and Universit\'e Joseph Fourier, Grenoble F-38042, France\\
$^{5}$ IMEP-LAHC, Grenoble INP, Minatec, Grenoble F-38016, France\\
$^{6}$ IEMN, Cit\'e scientifique, Villeneuve d'Ascq F-59652, France\\
$^{*}$ Correspondence to: (F.M.) frederico.rodrigues@uclouvain.be, (B.H.) benoit.hackens@uclouvain.be.}
\date{\today}

\begin{abstract}
The unique properties of quantum Hall devices arise from the ideal one-dimensional edge states that form in a two-dimensional electron system at high magnetic field. 
Tunnelling between edge states across a quantum point contact (QPC) has already revealed rich physics, like fractionally charged excitations, or chiral Luttinger liquid.
Thanks to scanning gate microscopy, we show that a single QPC can turn into an interferometer for specific potential landscapes. 
Spectroscopy, magnetic field and temperature dependences of electron transport reveal a quantitatively consistent interferometric behavior of the studied QPC.
To explain this unexpected behavior, we put forward a new model which relies on the presence of a quantum Hall island at the centre of the constriction as well as on different tunnelling paths surrounding the island, thereby creating a new type of interferometer. 
This work sets the ground for new device concepts based on coherent tunnelling.\\ 
\end{abstract}


\maketitle

Electron phase coherence is the cornerstone of quantum devices and computation~\cite{Beenakker_SSP91, Fisher_Science2009}. 
In that perspective, quantum Hall (QH) devices are particularly attractive in view of their large coherence times ~\cite{Roulleau_PRL2008}. 
Quantum Hall edge states (ES) formed by Landau levels (LL) crossing the Fermi energy near sample borders are ideal one-dimensional (1D) systems in which scattering vanishes exponentially at low temperature $T$~\cite{Beenakker_SSP91, Huckestein_RMP1995}. Edge state loops surrounding potential hills or wells, referred to as localized states or quantum Hall islands (QHIs), then form unique zero-dimensional (0D) systems~\cite{Ilani_Nat04}. The last few years witnessed great progresses in the transport spectroscopy of model QH localized states created by patterning quantum dots~\cite{Altimiras_NatPhys09} or antidots~\cite{Goldman_sc95,Goldman_PR98,Sim_pr2008} in a two-dimensional electron system (2DES).

\begin{figure}
\centering
\begin{center}
\includegraphics[width=10 cm]{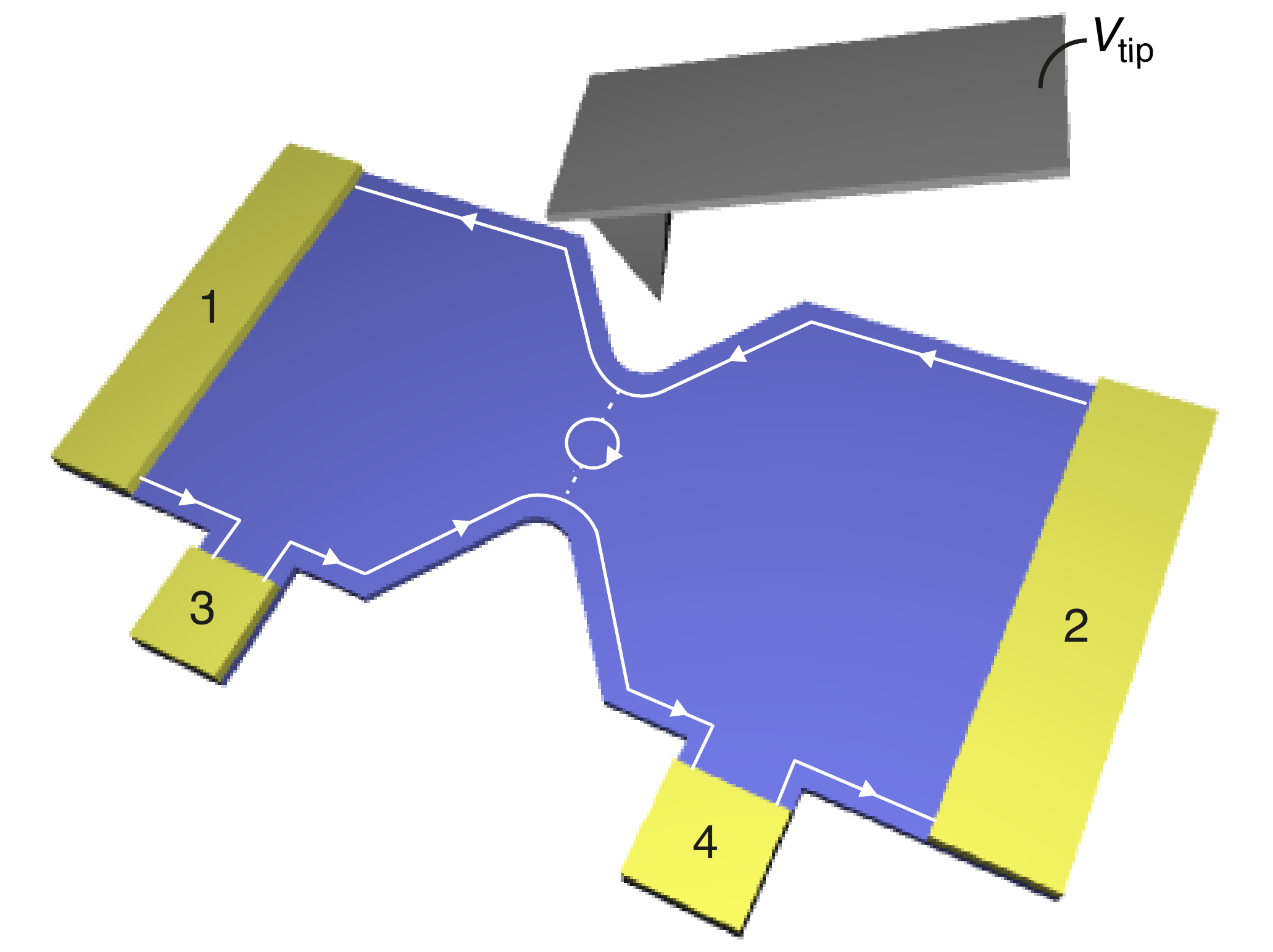}
\caption{{\bf Schematic representation of our model and experimental setup.} 
Tunnelling paths (dotted lines) connect opposite ES through a quantum Hall island (circle). Current-carrying contacts (1-2) and voltage probes (3-4) allow resistance measurements. (only one edge state is represented for the sake of clarity)
}
\label{fig:fig1}
\end{center}
\end{figure}

In parallel, new tools were developed to probe the microscopic structure of confined electron systems in the QH regime. In particular,  scanning gate microscopy~\cite{Topinka_Science2000,Crook_PRB2000,Pioda_PRL2004,Hackens_NatPhys2006,Martins_PRL2007,Pala_PRB2008,Pala_Nano2009} (SGM) makes use of a movable metallic tip, which is voltage-biased, to finely tune the electrons' confining potential in its vicinity. This way, the geometry of propagating edge states and localized states can be modified at will~\cite{Paradiso_PRB2011}.
Very recently, SGM allowed us to locate active QHIs in a QH interferometer~\cite{Hackens_NatComm2010}. 
Importantly, it appeared that QHIs do not only form around antidots, but potential inhomogeneities also induce QHIs in the arms or near the constrictions connecting a quantum ring to source and drain reservoirs~\cite{Hackens_NatComm2010}.
Therefore, lateral confinement, e.g. in Quantum Point Contacts (QPCs), offers the possibility to connect a QHI to ES through tunnel junctions, and thus form a new class of 1D-0D-1D QH devices (Fig.~\ref{fig:fig1}). In this case, the 0D island is characterized by a weak coupling ($\sigma<<e^2/h$) and a large charging energy ($E_c = e^2/C>>k_B T$) ($C$ is the island capacitance), which induce Coulomb blockade (CB)~\cite{Beenakker_SSP91}.
In such devices, Aharonov-Bohm (AB) like  oscillations of the resistance can be explained by Coulomb coupling between fully occupied LLs and confined states in the QHI~\cite{Taylor_PRL1992,Kataoka_prl1999,Sim_pr2008, Zhang_PRB09,Hackens_NatComm2010,Rosenow_PRL2007}.  
It was also suggested that AB oscillations reported on a QPC \cite{Loosdrecht_PRB88} could be attributed to tunnelling paths around the saddle point  \cite{Jain_PRB88}.
In contrast, transport through QH devices, but in the strong coupling limit ($\sigma>>e^2/h$), revealed coherent effects analog to those observed in optical Mach-Zehnder~\cite{Ji_Nature2003,Neder_PRL2006,Roulleau_PRL2008} or Fabry-P\'erot~\cite{Sivan_PRB1989, VanWees_PRL1989,McClure_prl2009,Zhang_PRB09,Yamauchi_PRB09,Ofek_PNAS10,Halperin_PRB2011} interferometers. 

Here, we examine an unexplored regime of transport across a QPC where QH edge states are weakly coupled, but phase coherence is preserved. 
The SGM tip is used as a nanogate to tune the potential landscape and hence edge states' pattern and coupling.
At first sight, one expects that transport should be driven by tunnelling, and possibly by Coulomb blockade if a quantum Hall island were mediating transport between edge states (Fig.~\ref{fig:fig1})~\cite{Hackens_NatComm2010}. Indeed, SGM and magnetoresistance data corroborate with Coulomb blockade across a QHI located near the saddle point of the QPC.
However, temperature dependence and scanning gate spectroscopy show clear signatures of quantum interferences.
Since, up to now, such interferences were exclusively observed in open QH devices, this observation sets the stage for a new electron transport scenario. We propose a new model that provides a quantitative interpretation of the data.

\section{Results}

\begin{figure}
\centering
\begin{center}
\includegraphics[width=10 cm]{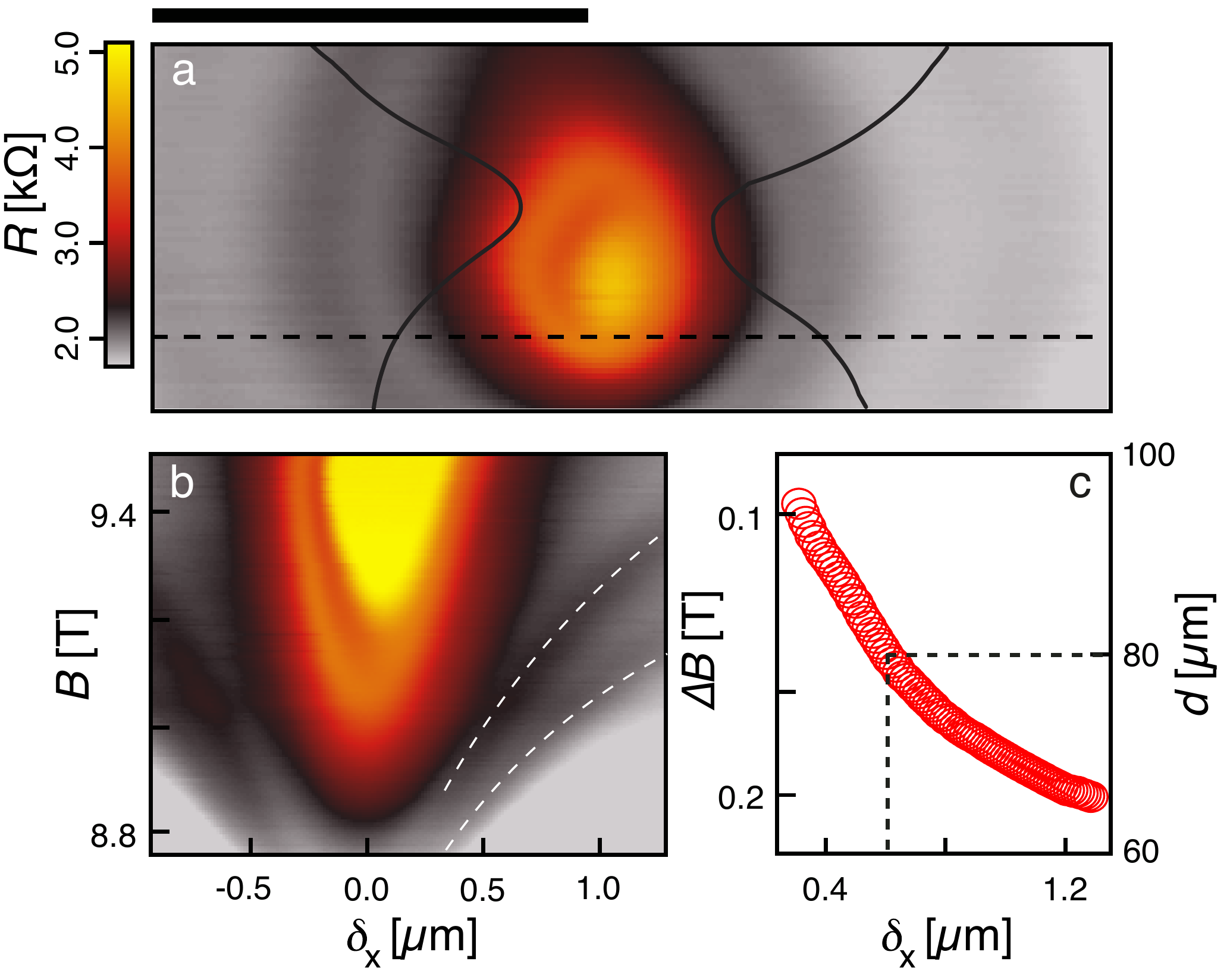}
\caption{{\bf Imaging tunnelling across a QPC.}
(a) SGM map at $B$~=~9.5~T,  $T$~=~4.2 K, and $V_{\mathrm{tip}}$~=~-4~V. Continuous lines correspond to the edges of the QPC. The black bar represents 1~$\mu$m.
(b) $B$-dependence of $R$-profiles over the region marked with a dashed line in (a), with $V_{\mathrm{tip}}$~=~-6~V.
Using Eq.~(\ref{rosenow}) for the two consecutive fringes highlighted with the white dashed lines in (b), we calculate in (c) the diameter of the QHI as the tip-QHI distance $\delta_\mathrm{x}$ is varied.
}
\label{fig:fig2}
\end{center}
\end{figure}

Our sample is a QPC etched in an InGaAs/InAlAs heterostructure holding a 2DES 25~nm below the surface. The QPC lithographic width is 300 nm. All the experiments were performed at temperature between 4.2 K and 100 mK, in a dilution refrigerator.
Here, the perpendicular magnetic field $B\sim9.5$~T, which corresponds to a LL filling factor $\nu \sim 6$ in the 2DES.
The SGM experiment is schematically depicted in Fig.~\ref{fig:fig1}. It consists in scanning a metallic atomic force microscope tip, polarized at voltage $V_{\mathrm{tip}}$, along a plane parallel to the 2DES at a tip-2DES distance of 50 nm while recording a map of the device resistance $R$~\cite{Hackens_NatPhys2006,Martins_PRL2007}. The QPC resistance is defined as $R=dV/dI$, where $V$ and $I$ are the voltage and the current through the device, respectively. 

The 2DES being on a quantized Hall plateau, whenever some current tunnels between opposite edge channels, $R$ deviates from the zero value expected in QH systems at very low $T$~\cite{Buttiker_PRB1988, Huckestein_RMP1995, Hackens_NatComm2010}. 
In our case, the SGM resistance map recorded at $B$~=~9.5~T, $V_{\mathrm{tip}}$~=~-4~V and $T$~=~4.2~K and presented in Fig.~\ref{fig:fig2}(a) reveals concentric fringes superimposed on a slowly varying background. The origin of the background, related to reflection of ES at the QPC, is discussed in the supplementary information.
The fringe pattern can easily be understood in the presence of a QHI surrounding a potential hill, close to the saddle point of the QPC and tunnel-coupled to the propagating ES (Fig.~\ref{fig:fig1}). Indeed, approaching the polarized  tip gradually changes the potential of the QHI, and hence its area $A$, defined as the surface enclosed by the "looping" ES. The enclosed magnetic flux $\phi$ varies accordingly and the tip generates iso-$\phi$ lines when circling around the QHI. Since adding one flux quantum $\phi_{\mathrm{0}}$ corresponds to trapping one electron per populated LL in the island, CB oscillations are generated whenever $B$ or $A$ are varied~\cite{Rosenow_PRL2007}, thereby producing AB-like oscillations~\cite{Taylor_PRL1992,Kataoka_prl1999,Sim_pr2008, Zhang_PRB09,Hackens_NatComm2010}.  Isoresistance lines visible on Fig.~\ref{fig:fig2}(a) are, therefore, iso-$\phi$ lines that are crossed as the tip-island distance is varied~\cite{Hackens_NatComm2010}. 
Consequently, the center of concentric fringes in Fig.~\ref{fig:fig2}(a) indicates the position of the active QHI, which connects opposite propagating edge channels through tunnel junctions (Fig.~\ref{fig:fig1}). 

In the framework of this model, the area of the QHI can be determined thanks to the $B$-dependence of AB-like oscillations~\cite{Rosenow_PRL2007}: 
\begin{equation}
	\Delta B= (\phi_{\mathrm{0}}/A)/N
\label{rosenow}
\end{equation}
where $N$ is the number of completely filled LL in the bulk (here $N=6$). 
The combined effect of moving the tip along the dashed line in Fig.~\ref{fig:fig2}(a) and changing $B$ is illustrated in Fig.~\ref{fig:fig2}(b) for $V_{\mathrm{tip}}$~=~-6~V. 
Along the $B$-axis, AB-like oscillations are highlighted with the white dashed lines. The negatively polarized tip approaching the QHI raises its potential, which increases its area $A$, and hence reduces the magnetic field that separates two resistance peaks $\Delta B$. This is illustrated in Fig.~\ref{fig:fig2}(c), where we assume that the QHI has a surface equivalent to that of a disk with diameter $d$ obtained from Eq.~(\ref{rosenow}): $d$ is found to increase from $\sim$65 nm to $\sim$95 nm as the tip-island distance $\delta_\mathrm{x}$ decreases from 1300 nm to 300 nm, respectively. 
Noteworthy, as expected for  Coulomb dominated transport in a QH interferometer, increasing $B$ is equivalent to applying a more negative $V_{\mathrm{tip}}$, yielding a positive $dV_{\mathrm{tip}}/dB$ for isoresistance stripes \cite{Zhang_PRB09,Ofek_PNAS10,Halperin_PRB2011}. Since approaching the negatively charged tip has the same effect as decreasing $V_{\mathrm{tip}}$, Fig.~\ref{fig:fig2}(b) seems consistent with the  Coulomb dominated transport. 

\begin{figure}
\centering
\begin{center}
\includegraphics[width=10cm]{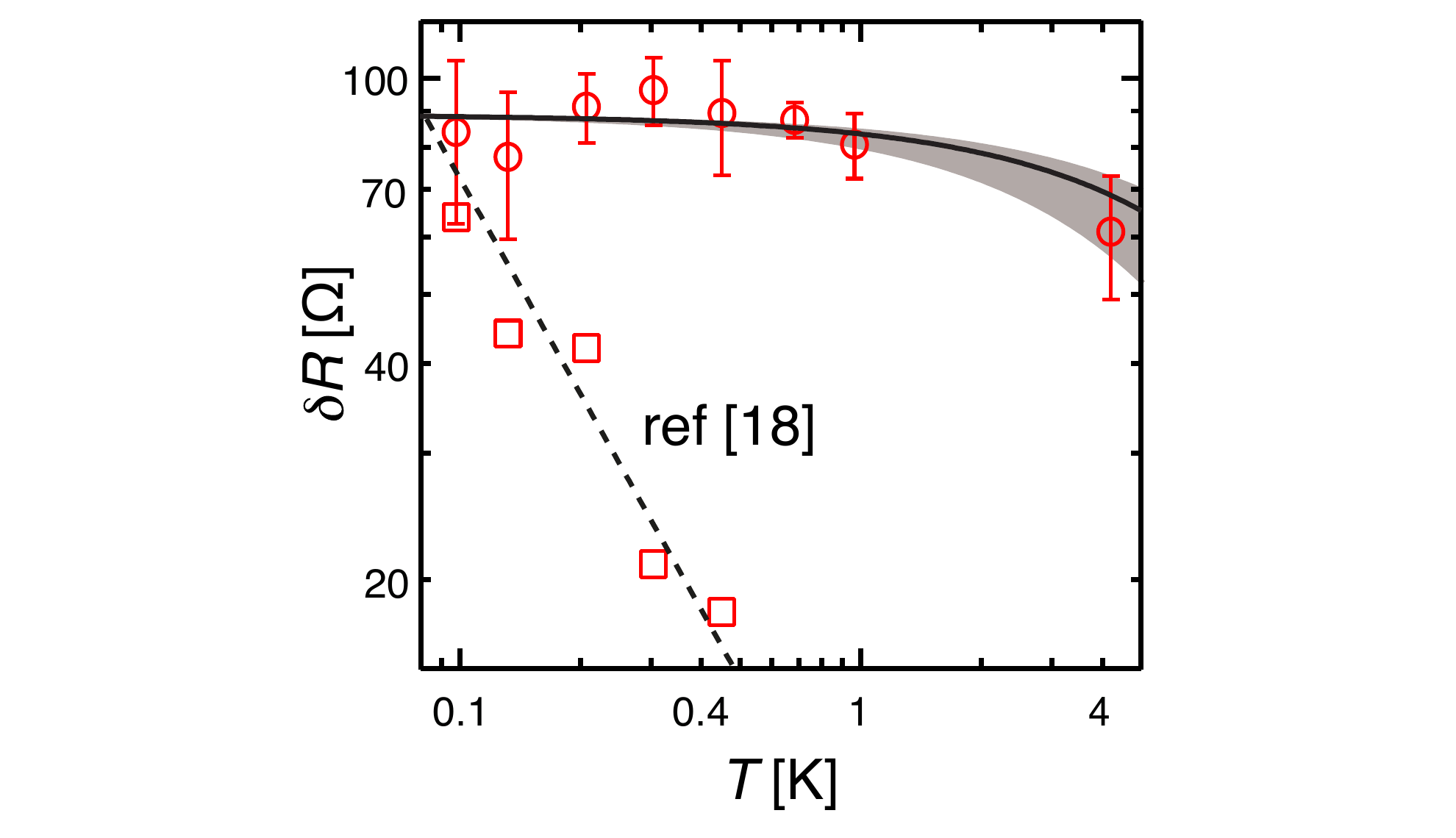}
\caption{{\bf Temperature dependence : Coulomb blockade vs coherent transport.}
$\delta R$ vs $T$ obtained from SGM maps with $V_{\mathrm{tip}}$~=~-1~V (circles) and from data in ref.~\cite{Hackens_NatComm2010} (squares). 
The dashed line corresponds to a $T^{-1}$ dependence. The gray region corresponds to an exponential dependence exp$(-T/ T_{\mathrm{0}})$ with $9.5~\mathrm{K}<T_{\mathrm{0}}<19.1~\mathrm{K}$,  consistent with magnetoresistance data and edge state velocity estimate along Ref.~\cite{McClure_prl2009}. The solid line corresponds to $T_{\mathrm{0}} = $~16.2~K, consistent with the spectroscopy data (see text).
}
\label{fig:fig3}
\end{center}
\end{figure}

But, surprisingly, the temperature dependence of fringes amplitude ($\delta R$, measured on SGM maps), shown on Fig.~\ref{fig:fig3}, reveals a peculiar behaviour: it clearly does not follow the $T^{-1}$ dependence expected in the quantum regime of CB~\cite{Kouwenhoven_AS1997,Yacoby_PRL1995,Hackens_NatComm2010} (data from ref.~\cite{Hackens_NatComm2010} are presented for comparison in Fig.~\ref{fig:fig3}).
Instead, $\delta R$ deceases very slowly from 100 mK to 4.2 K.
Indeed, for coherent transport through a Fabry-P\'erot geometry, thermal smearing of interference gives rise to a temperature dependence $\delta R(T) \sim$exp$(-T/ T_{\mathrm{0}})$ in the low temperature regime. In contrast, for transport processes involving a weakly coupled Coulomb island,  this form for $\delta R(T)$ is expected only for  temperatures larger than the charging energy ~\cite{Halperin_PRB2011}.
In the Fabry--P\'erot situation, $T_{\mathrm{0}}$ is linked to the excited states level spacing $\Delta E_{\mathrm{Ex}}$ according to the relation $T_{\mathrm{0}} = \Delta E_{\mathrm{Ex}}/k_{\mathrm{B}} = 2\hbar v/(d k_{\mathrm{B}})$ where $v$ is the local edge state velocity, related to the gradient of the confining potential.
From experimental data measured in a GaAs QH Fabry-P{\'e}rot interferometer~\cite{McClure_prl2009}, one can infer that, in our sample, $5 \times 10^{4}~\mathrm{m/s}<v< 10^{5}$ m/s.
Given this range for $v$, and $d \sim$ 80 nm (from Fig.~\ref{fig:fig2}(c), taking into account that the $T$-dependence data were measured at $\delta_\mathrm{x} \sim$ 630 nm), we obtain the range of $T$-dependence represented as a gray region in Fig.~\ref{fig:fig3}, which reproduces quite well the behaviour observed experimentally. The corresponding range of  $9.5~\mathrm{K} <T_{\mathrm{0}}<19.1$~K is consistent with the low temperature limit and hence with a Fabry-P\'erot behaviour.
Earlier experiments already evidenced such an exponential decay with temperature, but only in Mach-Zehnder and ballistic devices, which are known to be coherent~\cite{Yamauchi_PRB09,Roulleau_PRB2007,Litvin_PRB2008}.

\begin{figure}
\centering
\begin{center}
\includegraphics[width=10 cm]{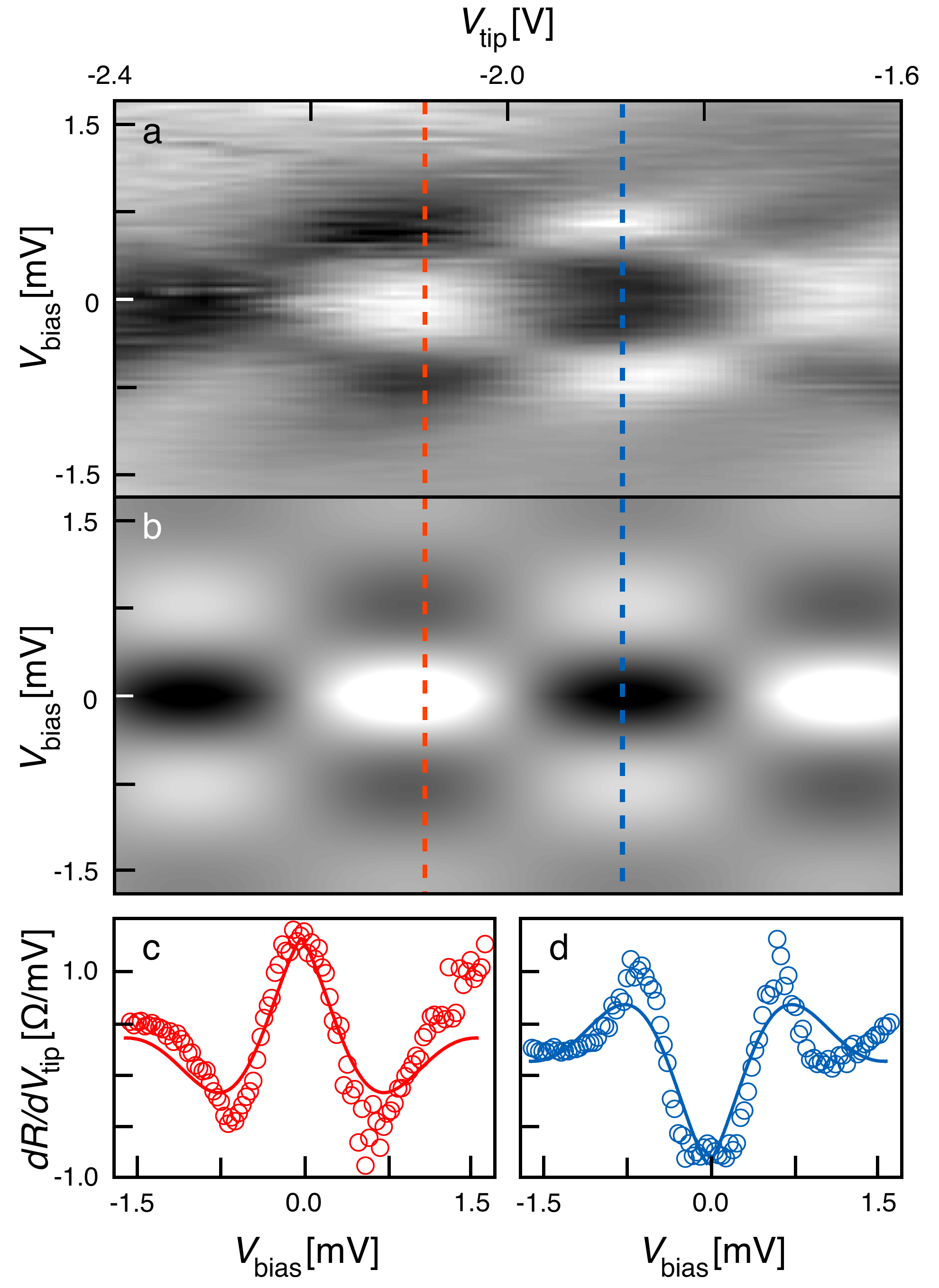}
\caption{ {\bf Evidence for coherent transport in spectroscopy\bf.}
(a) $dR/dV_{\mathrm{tip}}$ as a function of the dc component of $V_{\mathrm{tip}}$ and $V_{\mathrm{bias}}$ at $B=9.5$ T and $T$=100 mK. Voltage modulation of $V_{\mathrm{tip}}$ was set to 50~mV. 
(b) 2D fit of $dR/dV_{\mathrm{tip}}$ using Eq.~(\ref{checkerboard}). (c-d) Transresistance vs $V_{\mathrm{bias}}$ taken along the red (c) and blue (d) dashed lines in (a-b). The circles correspond to the experimental data and the continuous lines to the fit.
}
\label{fig:fig4}
\end{center}
\end{figure}

However, our main observation confirming the preserved electron phase coherence emerges from the analysis of non-linear transport through the QPC. 
Scanning gate spectroscopy is realized by positioning the tip right above the QHI, and sweeping both $V_{\mathrm{tip}}$ and the dc current $I$ through the QPC.
The voltage across our tunnel device, i.e. between propagating edge states, is the Hall voltage: $V_{\mathrm{bias}}=h/(e^2 N^*)I$~\cite{Buttiker_PRB1988}, where $N^*$ is the (integer) number of transmitted ES at the QPC (Fig.~S1 - supplementary material). 
The measurement configuration is indeed identical to that used to perform conventional electrical spectroscopy on isolated quantum dots. If the QHI were weakly tunnel coupled to the propagating edge states, one would expect to observe a "Coulomb diamond" pattern~\cite{Martins_NJP2013}. 
Fig.~\ref{fig:fig4}(a) shows $dR/dV_{\mathrm{tip}}$ as a function of the dc component of $V_{\mathrm{tip}}$ and $V_{\mathrm{bias}}$. 
Instead of Coulomb diamonds, the spectroscopy displays a checkerboard pattern 
of maxima and minima, indicating that both $V_{\mathrm{tip}}$ and $V_{\mathrm{bias}}$ tune the interference of transiting electrons.
Each bias independently adds a phase shift between interfering paths, so that the transresistance is modulated by a product of cosines and an exponential term accounting for a voltage-dependent dephasing induced by electrons injected at an energy $e|V_{\mathrm{bias}}|$ ~\cite{McClure_prl2009,Yamauchi_PRB09,Roulleau_PRB2007,Wiel_PRB2003}:
\begin{equation}
 	\frac{dR}{dV_{\mathrm{tip}}} = D \cos \left( 2 \pi \frac{V_{\mathrm{bias}} }{ \Delta V_{\mathrm{bias}}} \right)
	 \cos \left( 2 \pi \frac{V_{\mathrm{tip}} }{\Delta V_{\mathrm{tip}}} +\varphi \right) \exp \left(- 2 \pi \gamma \left(\frac{V_{\mathrm{bias}}}{\Delta V_{\mathrm{bias}}}\right)^n\right),
	 \label{checkerboard}
\end{equation}
where $D$ is the zero-bias visibility of the oscillations, $\Delta V_{\mathrm{tip}}$ is the oscillation period induced by $V_{\mathrm{tip}}$, $\varphi$ is a constant phase factor,
$\Delta V_{\mathrm{bias}}=4\hbar~v/(ed)$ is the oscillation period along the $V_{\mathrm{bias}}$ axis, and  $\gamma$ is directly related to the $V_{\mathrm{bias}}$-dependent dephasing rate: $\tau_{\varphi}^{-1}=\gamma(e|V_{\mathrm{bias}}|)/2\hbar$~\cite{McClure_prl2009}. $n$ varies from 1 to 2 according to Ref.~\cite{McClure_prl2009,Yamauchi_PRB09,Roulleau_PRB2007,Wiel_PRB2003} and was set to 1 as we could not discriminate from fitting the data. As shown in Figs.~\ref{fig:fig4}(b-d), we obtain an excellent fit of the data in Fig.~\ref{fig:fig4}(a) using Eq.~(\ref{checkerboard}) with with a transist time $\tau_t ~= d/v =~1.7\times10^{-12}$~s, and a parameter $\gamma$ = 0.2 in the range found in Ref.~\cite{McClure_prl2009}. Note that in such a small QHI, $\tau_t$ turns out to be smaller by at least one order of magnitude than the intrinsic $\tau_{\varphi}$ in the same 2DES~\cite{Hackens_PRL2005}. This renders coherent resonant tunnelling through the whole QHI device possible.

\section{Discussion}
To interpret $\Delta V_{\mathrm{tip}}$ obtained from the fit, one first notes that $R$ evolves very similarly when changing either $V_{\mathrm{tip}}$ or $B$ in the vicinity of $B =$ 9 T (Fig. S2 - supplementary material). 
Therefore, one can convert $\Delta V_{\mathrm{tip}}$ into an equivalent $\Delta B$, through a lever arm $\Delta B/\Delta V_{\mathrm{tip}}=$ 0.108 T/V. 
Hence, $\Delta V_{\mathrm{tip}} = 0.46$ V corresponds to $\Delta B$ = 50 mT for the AB-like oscillations. 
In that range of $V_{\mathrm{tip}}$, $N^{*}=5$ (Fig.~S1(d)).
This means that $d=2 \sqrt{\phi_{0}/(\pi N^{*} \Delta B)}=145$ nm, consistent with data in Fig.~\ref{fig:fig2}(c) since $d$ is at a maximum when the tip is above the QHI ($\delta_\mathrm{x}=0$). 
Moreover, given the value of $\tau_t $~=~$1.7\times10^{-12}$~s found in fitting the spectroscopy data, one obtains $v = 8.5 \times 10^4$ m/s, within the range of values that was expected from data in ref.~\cite{McClure_prl2009}, and in agreement with the exponential temperature dependence in Fig.~\ref{fig:fig3}.
We therefore have a fully consistent picture that explains all magnetoresistance, temperature dependence and spectroscopy data, and shows that tunnelling across the QHI is indeed coherent.

One fundamental question remains: why do we observe two distinct behaviours of transport through apparently similar QH devices, Coulomb blockaded transport in our previous work~\cite{Hackens_NatComm2010}, and coherent transport in this one? 
The qualitative difference cannot be explained by the fact that $d$ is smaller than previously examined QHIs. $T_{\mathrm{0}}$ and $\Delta V_{\mathrm{bias}}$ would be reduced proportionally, but not enough to explain the observed $T$-dependence and spectroscopy. On  the other hand, signs of coherent transport through CB quantum dots were only obtained for symmetric tunnel junctions~\cite{Yacoby_PRL1995} that allow resonant tunnelling instead of sequential tunnelling. In that framework, one might thus ascribe the loss of electron coherence in other QHIs to an asymmetry of tunnel junctions.
However, a difference in the transmission coefficients $T_{\mathrm{c}}$ of the tunnel barriers may point towards an alternative explanation. In the coherent regime, we find a rather strong coupling between the QHI and propagating ES ($0.27<T_{\mathrm{c}}<0.43$), which contrasts with the Coulomb blockade regime where $T_{\mathrm{c}}<<1$~\cite{Houten_1992}.
A similar trend is observed in transport experiments at $B$ = 0 T on carbon nanotubes~\cite{Biercuk2008}: phase coherence is maintained when electrons tunnel through barriers with a large transmission coefficient, so that interference effects can be observed.

\begin{figure}
\centering
\begin{center}
\includegraphics[width=10 cm]{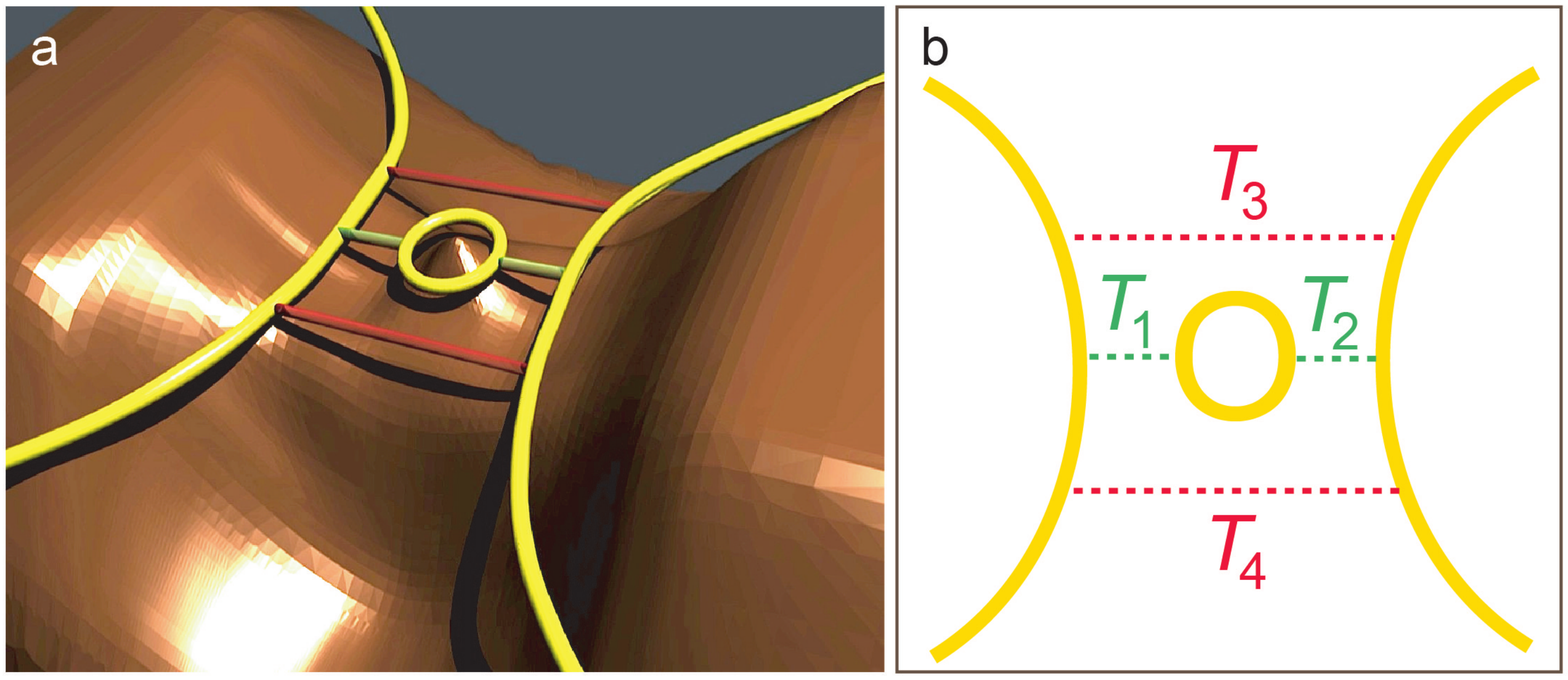}
\caption{{\bf  Potential landscape and tunnelling paths across the QPC.}
(a) Schematic representation of the electrostatic potential in the vicinity of the QPC (in brown), with the ES in yellow and the tunnelling paths connecting opposite ES (red and green). Only one edge state is represented, for the sake of clarity. 
(b) top view of the three-dimensional figure in (a), with the various tunnelling probabilities $T_{i}$ between edge states.
}
\label{fig:fig5}
\end{center}
\end{figure}

Up to this point, our analysis is based on the presence of a QHI near the QPC, connected to propagating ES on both sides through tunnelling paths (Fig.~\ref{fig:fig1}). However, one could imagine the presence of additional tunnelling paths between propagating ES, in the vicinity of the QPC saddle point. The resulting model is presented in Fig.~\ref{fig:fig5}(a-b). 
While the "green" paths occur naturally when propagating ES and the QHI are close enough, the "red" paths may originate from potential anharmonicities (\emph{i.e.} non parabolicity) on both sides of the saddle point, similar to the fast potential variations suggested in Ref.~\cite{Jain_PRB88}.
In this model, transport depends in principle on the various tunnelling probabilities, denoted $T_{1,2}$  and $T_{3,4}$ in Fig.~\ref{fig:fig5}. However, the presence of the QHI should always induce oscillations in the magnetoresistance and spectroscopy of the QPC, either because it is enclosed in an interferometer, created by the "red" paths and propagating ES, when $T_{1,2}<T_{3,4}$, or because tunnelling occurs directly through it ($T_{1,2}>T_{3,4}$) as discussed above (Fig.~\ref{fig:fig1}). 
Therefore, whichever  $T_{1,2}$ or $T_{3,4}$ dominates, transport is still controlled by the flux trapped in the QHI and hence its Coulomb charging, so that the analysis developed above to extract parameters from the magnetoresistance and spectroscopy are still valid. In that case, i.e. $T_{3,4}>T_{1,2}$, the amplitude of the fringes leads us to $0.043<T_{3,4}<0.078$ (for details see Supplementary Information). 

%Noteworthy, SGM would be well suited to tune tunnelling probabilities and discriminate which of $\Gamma_{1,2}$ or $\Gamma_{3,4}$ are the effective tunnelling paths.
In summary, we report first evidence for preserved electron phase coherence in tunnelling across a quantum point contact in the quantum Hall regime. We propose a framework that explains all magnetoresistance, temperature dependence and spectroscopy data. 
This scenario relies on the presence of a potential hill that generates a quantum Hall island near the saddle point of the QPC.
Our data therefore provide new signatures of coherent tunnelling in an ultra-small QH device.\\

\section{Methods}
{\bf Device fabrication and 2DES parameters.} Our device is fabricated from a InGaAs/InAlAs heterostructure grown by molecular beam epitaxy where a  2DES is confined 25 nm below the surface (the layer sequence of this heterostructure is detailed in \cite{Hackens_NatPhys2006,Martins_PRL2007}). 
The QPC was patterned using e-beam lithography followed by wet etching. 
The QPC resistance $R$ is measured in a four-probe configuration: a low-frequency (typically 10 to 20 Hz) oscillating current $I$ is driven between contacts 1 and 2 on Fig.~\ref{fig:fig1}, and $V$ is measured between contacts 3 and 4 using a lock-in technique, with $V$ across the QPC always less than $k_{B}T/e$. 
Next to the QPC, we patterned a Hall bar where we measured a low-$T$ electron density and mobility of $1.4 \times 10^{16}~\mathrm{m^{-2}}$ and $4~\mathrm{m^{2}/Vs}$, respectively. 

{\bf SGM and SGS techniques.} All the experiments are carried out inside a $^3$He/$^4$He dilution refrigerator where a home-made cryogenic atomic force microscope (AFM) was integrated~\cite{Hackens_NatComm2010}. The AFM is based on a quartz tuning fork to which a commercial metallized Si cantilever (model CSC17 from MikroMasch) is glued by means of a conductive silver epoxy.  
We image the sample topography by imposing a feedback loop on the shift in the tuning fork resonant frequency and using standard dynamic AFM mode of operation.
After locating the QPC we perform SGM. 
It consists of scanning the tip along a plane parallel to the 2DES at constant distance of 25~nm from the surface, i.e. 50~nm from the 2DES, with a bias voltage $V_{\mathrm{tip}}$ applied to the tip and recording simultaneously the device resistance $R$.
At the end of a set of SGM experiments, we image the topography of the QPC to ensure that, during that period, the position of the QPC did not change.
The SGS is performed by positioning the AFM tip at a fixed position in the vicinity of the QHI and by adding a dc current $I$ to the lock-in ac signal between contacts 1 and 2 (Fig.~\ref{fig:fig1}). 
The voltage between edge states $V_{\mathrm{bias}}$ is obtained by multiplying the dc current $I$ by $h/(e^2 N^*)$.
The transresistance $dR/dV_{\mathrm{tip}}$ is measured with a second lock-in using an ac modulation of $V_{\mathrm{tip}}$.

{\bf Acknowledgements:}\\
The authors are grateful to M. Treffkorn and T. Hyart for helpful discussions.
F.M. and B.H. are postdoctoral and associate researchers with the Belgian FRS-FNRS, respectively. This work has been supported by FRFC grants no. 2.4.546.08.F and 2.4503.12, FNRS grant no. 1.5.044.07.F, by the FSR and ARC program "Stresstronics", by BELSPO (Interuniversity Attraction Pole IAP-6/42), and by the PNANO 2007 program of the ANR (MICATEC project). V.B. acknowledges the award of a Chaire d'excellence by the Nanoscience Foundation in Grenoble.

{\bf Author contributions:}\\
F. M., B. H. and S. F. performed the low-temperature SGM experiment; F. M., B. H., V. B. and B. R. analysed the experimental data; L. D. and X. W. grew the InGaAs heterostructure; B. H. processed the sample; B. H., S. F. and F. M. built the low temperature scanning gate microscope; B. H., F. M., S. F., H. S., S. H., M. P. and V. B. contributed to the conception of the experiment; F.M., B. H. and V. B. wrote the paper and all authors discussed the results and commented on the manuscript.

{\bf Additional information:}\\
Supplementary Information accompanies this paper.\\
Competing financial interests: The authors declare no competing financial interests.

\bibliographystyle{prsty}

\pagebreak

\section {Supplementary Information for:
``Coherent tunnelling across a quantum point contact in the quantum Hall regime''}

\section{Origin of the background in SGM}

\begin{figure}
\renewcommand{\figurename}{FIG. S}
\begin{center}
\includegraphics[width=120 mm]{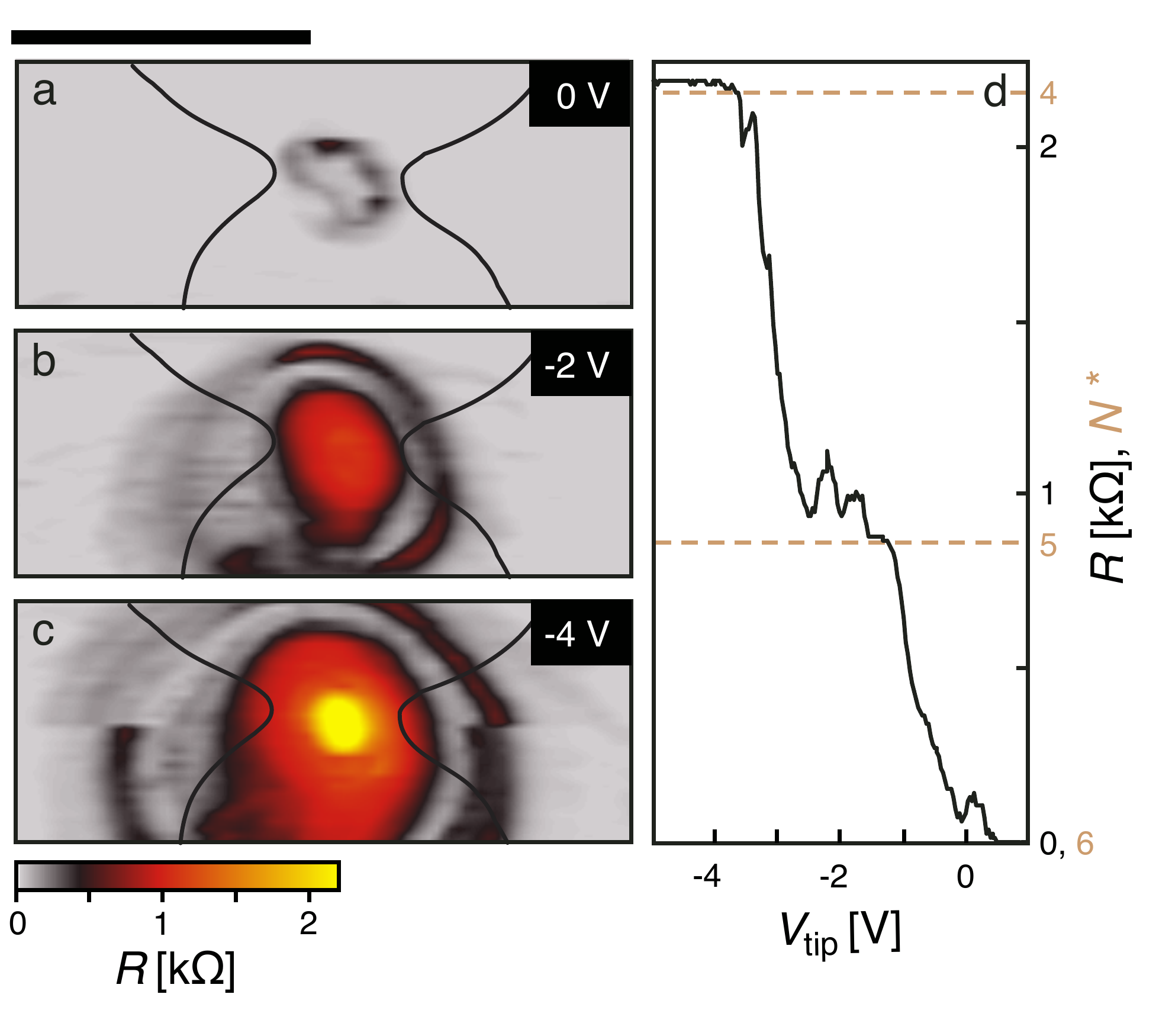}
\caption{
(a-c) Consecutive SGM images obtained at $T = 100$~mK, $B = 9.5$~T and $V_{\rm tip}$~=~0, -2 and -4~V, respectively. 
The top black bar represents 1~$\mu$m.
(d) $R$ vs $V_{\rm tip}$ with the tip positioned near the saddle point of the QPC. The brown dashed lines indicate the resistance expected for $N^*$.
}
\label{fig:Sup_1}
\end{center}
\end{figure}

Fig. S\ref{fig:Sup_1} allows inferring the origin of the broad background in SGM images. 
By sweeping $V_{\mathrm{tip}}$ below 0 V, the SGM maps measured at 100 mK shown in Figs. S\ref{fig:Sup_1}(a-c) reveal concentric fringes marking the presence of a QHI near the saddle point of the QPC. 
The diameter of circling fringes increases with decreasing $V_{\mathrm{tip}}$, consistent with the observations in Fig.~S~\ref{fig:Sup_1}. 
Importantly, the SGM pattern around the QPC exhibits a strong variation that adds to the concentric fringes on Figs.~S\ref{fig:Sup_1}(b-c). 
The origin of the stronger contrast is found by positioning the tip near the saddle point of the QPC and continuously decreasing $V_{\mathrm{tip}}$ (Fig. S\ref{fig:Sup_1}(d)). 
The main trend shows step-like increases of the device resistance which can be understood by invoking ES reflections at the constriction. Decreasing $V_{\mathrm{tip}}$ raises the energy of the saddle point and decreases the local filling factor $\nu^*$ near the constriction. Every time $\nu^*$ passes a half integer value,  an ES is totally reflected and the device resistance shifts to the next  plateau given by $R=h/ e^2 (1/N^*-1/ N)$~\cite{Aoki_PRB2005, Buttiker_PRB1988} (brown dashed lines in Fig. S\ref{fig:Sup_1}(d)), where $N^*$ is the (integer) number of transmitted ES at the constriction. 
The presence of oscillations superimposed on the first plateau at $N^*$~=~5, similar to those around $V_{\mathrm{tip}}$~=~0~V, indicate that the QHI is active even when one ES is reflected. 

\section{Calculation of transmission coefficients across a QHI}

Here we summarize the details concerning the determination of the coefficients of transmission across the tunnel barriers defining the QHI in the coherent regime (we assume here that all barriers have equal transmission coefficients).
In this work it was found that the peak-to-peak amplitudes in the coherent regime ($\Delta R_{\mathrm{}}$) were within the following intervals: $170 \Omega$ $(N^*=5)$ $< \Delta R_{\mathrm{}} <200\Omega$ $(N^*=6)$.
Assuming that $\Delta R= h/e^2(1/(N^*-T_{\mathrm{t}})-1/N^*)$ ~\cite{Aoki_PRB2005, Buttiker_PRB1988} where $T_{\mathrm{t}}$ is the total transmission through the QHI (i.e. taking into account the two barriers defining the QHI), we conclude that $T_{\mathrm{t}}$ is the interval: $0.16 <T_{\mathrm{t}}< 0.27$.

%\begin{itemize}
%\item $0.149 < {T_{\mathrm{t}}}^{\mathrm{cd}}< 0.228$ in the Coulomb dominated regime;
%\item $0.159 <{T_{\mathrm{t}}}^{\mathrm{c}}< 0.267$ in the coherent regime.
%\end{itemize}

In Fig.~S~\ref{fig:Sup_3} we draw the two models considered in the main article. In the following subsections we deduce the coefficients of transmission of the tunnel barriers for the two different situations presented in Fig.~S~\ref{fig:Sup_3}(a) and (b).

%\subsection{QHI at the centre of a QPC in the Coulomb dominated regime}
%
%We consider the situation represented in Fig.~S~\ref{fig:Sup_3}(a) in the Coulomb dominated regime, i.e. when the fringes amplitude follow a $T^{-1}$ temperature dependence. In this case the total transmission is given by \cite{Houten_1992}:
%\begin{equation}
%	{T_{\mathrm{t}}}^{\mathrm{cd}} = \frac{h}{8k_{\mathrm{B}}T} \frac{\Gamma_{\mathrm{1}}\Gamma_{\mathrm{2}}}{\Gamma_{\mathrm{1}}+\Gamma_{\mathrm{2}}} 
%	 \label{Coulomb_Dominated}
%\end{equation}
%where $k_{\mathrm{B}}$ is Boltzmann constant and $\Gamma_{\mathrm{1}}$ and $\Gamma_{\mathrm{2}}$ are the tunnelling rates (in Hz) through the barriers with the transmission $T_1$ and $T_2$, respectively. Assuming that $\Gamma_{\mathrm{1}}= \Gamma_{\mathrm{2}}=\Gamma$ and  $T_1= T_2=T_{\mathrm{c}}$, we have:
%\begin{equation}
%	\Gamma = \frac{16k_{\mathrm{B}}T}{h}{T_{\mathrm{t}}}^{\mathrm{cd}} 
%	 \label{Coulomb_Dominated}
%\end{equation}
%
%The transmission is, thereafter, obtained by dividing $\Gamma$ by the attempt frequency $\nu = E_{\mathrm{F}}/h$, where $E_{\mathrm{F}} = 76$~meV is the Fermi energy measured in our experiment. In this case we write:
%\begin{equation}
%	T_{\mathrm{c}} = \frac{16k_{\mathrm{B}}T}{E_{\mathrm{F}}} {T_{\mathrm{t}}}^{\mathrm{cd}} 
%	 \label{Coulomb_Dominated_2}
%\end{equation}
%and we find $2.7\times 10^{-4}<T_{\mathrm{c}}<4.1\times 10^{-4}$.

%-------------------------------
\subsection{QHI at the centre of a QPC}

We first consider the situation represented in Fig.~S~\ref{fig:Sup_3}(a) where a QHI is located at the centre of a QPC. 
In this case we assume that coherence is maintained during the multiple reflections. For an off-resonance condition and assuming $T_1= T_2=T_{\mathrm{c}}$,  $T_{\mathrm{c}}$ is given by: $T_{\mathrm{c}}=2{T_{\mathrm{t}}}(1+{T_{\mathrm{t}}})$ \cite{Datta_1995}, which implies that $0.27<T_{\mathrm{c}}<0.42$.

%-------------------------------
\subsection{Interferometer formed around the QPC saddle point}

In the case of an interferometer formed around the saddle point of a QPC, as illustrated in Fig.~S~\ref{fig:Sup_3}(b), we compute the reflection coefficient $1-{T_{\mathrm{t}}}$, taking into account interferences between different semiclassical paths for electrons : 1) a direct path along the edge state, which does not include transmission through the tunnel barriers, and 2) paths including multiple transmissions through the tunnel barriers. The reflection coefficient is then given by  \cite{Datta_1995}:

\begin{equation}
	1-{T_{\mathrm{t}}}= \frac{(1-T_3) (1-T_4)}{1 + T_3  T_4 - 2 \sqrt{T_3 T_4} cos(\varpi)}
	 \label{Coherent}
\end{equation}
where $\varpi$ is the phase difference accumulated along the two types of trajectories and $T_3$ and $T_4$ are the transmissions at each side of the saddle point. Assuming that $T_3 = T_4= T_{3,4}$, we obtain:  $0.043<T_{3,4}<0.078$.

\pagebreak

\begin{figure}
\renewcommand{\figurename}{FIG. S}
\begin{center}
\includegraphics[width=100 mm]{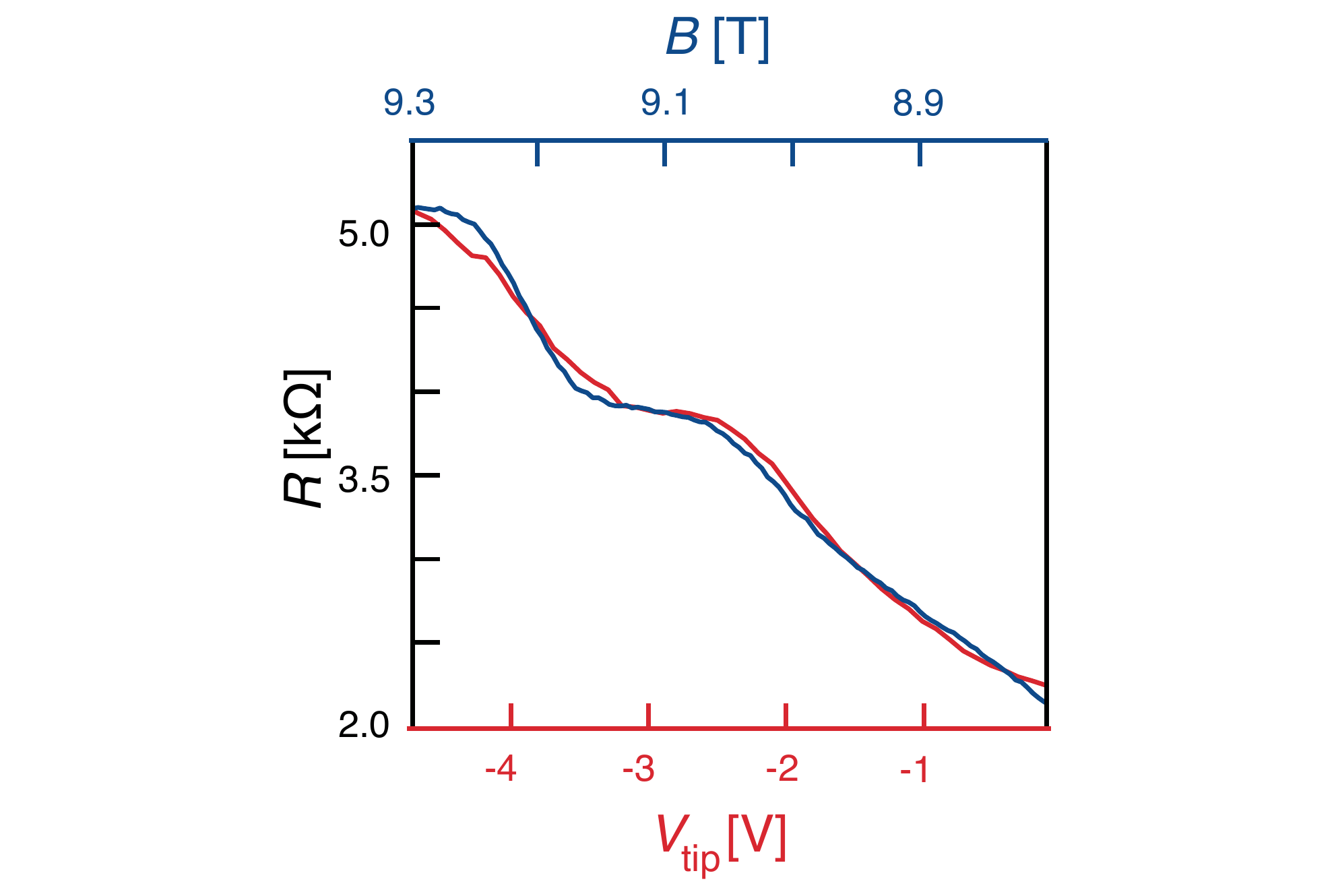}
\caption{QPC resistance vs $B$ (top axis) and $V_{\mathrm{tip}}$ (bottom axis) at $T = 4.2$~K.
}
\label{fig:Sup_2}
\end{center}
\end{figure}

\begin{figure}
\renewcommand{\figurename}{FIG. S}
\begin{center}
\includegraphics[width=120 mm]{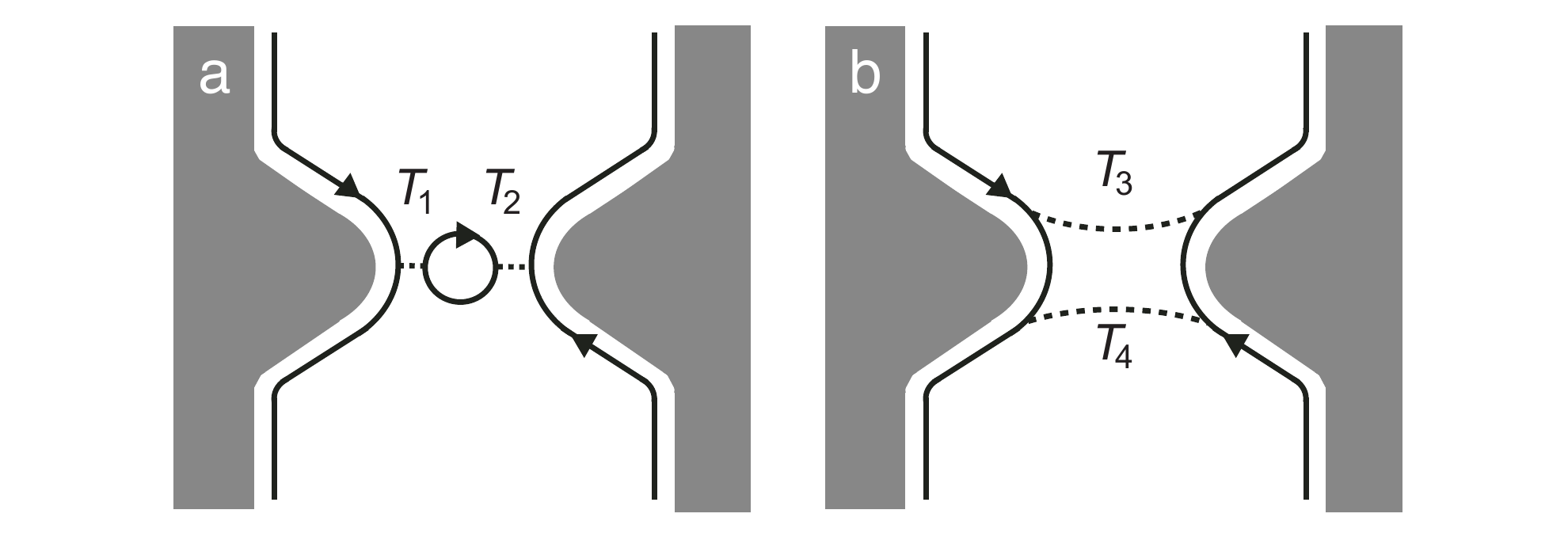}
\caption{ (a) Schematic representation of a QHI at the center of the QPC. 
Tunnelling paths (dotted lines) connect opposite ES through the QHI. $T_1$ and $T_2$ are transmission coefficients of the tunnel barriers between ES and the QHI.. 
(b) Alternative model for the situation at the QPC:  two tunnelling paths (dotted lines) on both sides of the saddle point
 connect counterpropagating edge states and form a closed loop. Note that in both cases, only one edge state is represented, for the sake of clarity
}
\label{fig:Sup_3}
\end{center}
\end{figure}

\end{document}